\documentclass[11pt]{article}

\usepackage{amsmath}
\usepackage{amssymb}
\usepackage{amsfonts}
\usepackage{amsthm}
\usepackage{bm}
\usepackage{graphicx}
\usepackage{subcaption}
\usepackage{multirow}
\usepackage{authblk}
\usepackage{natbib}
\usepackage{algorithm}
\usepackage{algorithmic}
\usepackage{dsfont}

\usepackage{tikz}
\usetikzlibrary{arrows,automata}
\usetikzlibrary{shapes.geometric, arrows}
\usetikzlibrary{positioning}

\begin{document}

\title{Influence Maximization With Deactivation In Social Networks}

\author[1]{K\"{u}bra Tan{\i}nm{\i}\c{s}}
\author[1]{Necati Aras}
\author[1]{\.{I}.K. Alt{\i}nel}

\affil[1]{Department of Industrial Engineering, Bo\u{g}azi\c{c}i University, \.{I}stanbul, Turkey}
\affil[ ] {\{kubra.taninmis,arasn,altinel\}@boun.edu.tr}

\date{}

\maketitle

\begin{abstract}
In this paper we consider an extension of the well-known Influence Maximization Problem in a social network which deals with finding a set of $k$ nodes to initiate a diffusion process so that the total number of influenced nodes at the end of the process is maximized. The extension focuses on a competitive variant where two decision makers are involved. The first one, the leader, tries to maximize the total influence spread by selecting the most influential nodes and the second one, the follower, tries to minimize it by deactivating some of these nodes. The formulated bilevel model is solved by complete enumeration for small-sized instances and by a matheuristic for large-sized instances. In both cases, the lower level problem, which is a stochastic optimization problem, is approximated via the Sample Average Approximation method.


\end{abstract}

\section{Introduction}
Social network analysis studies the behaviors of the actors in a social group consisting of “individuals”, “organizations” or “communities”, and the relationships between these entities. The relationships can be defined in various contexts such as economic, social, political, and analyzing the structure of these relationships helps to answer questions in different disciplines \citep{wasserman1994social}. Although this research area has been around for decades, it has attracted much more interest along with the growth in the number and prevalence of online social networking sites such as Facebook and Twitter \citep{scott2012social}. One of the issues of primary interest in social network analysis is social influence. It is based on the fact that one's ideas can affect those of his friends (actors it is linked to). Existence of social influence may result in the spread of ideas or information like an epidemic. Viral marketing is one of the widespread applications of the influence spread through a social network. Ideas or new technologies can quickly reach the masses, when influential individuals are targeted initially. The first study on the maximization of influence belongs to \citet{Domingos}. They claim that considering customers as independent entities causes suboptimal marketing decisions. \citet{Kempe} define the Influence Maximization Problem (IMP) as finding a set of $k$ nodes to start a diffusion so that the expected number of influenced nodes at the end of the diffusion is maximized.

The IMP is a stochastic optimization problem due to the uncertainties in the underlying diffusion model. The most common diffusion models in the IMP literature are Linear Threshold (LT), Independent Cascade (IC) and their extensions \citep{Kempe}. IMP is $\mathcal{NP}$-hard under both models, and its objective function is submodular. Greedy algorithms provide approximation guarantees for problems with monotone submodular objective function \citep{cornuejols1977exceptional,nemhauser1978analysis}. Using this property, \citet{Kempe} propose a greedy algorithm with a $(1-1/e-\epsilon)$ approximation ratio, where $e$ is Euler's number and $\epsilon$ is an arbitrarily small positive number. There are some other works that develop methods with the same ratio, such as \citet{Leskovec} and \citet{ChenWang}. Due to the high time-complexity of these methods, some research focuses on the scalability issue \citep{ChenYuan,wang2012scalable,tang2014influence}. \citet{Guney2017} formulates the IMP as stochastic program and approximates the optimal objective by the Sample Average Approximation method. \citet{wu2018two} propose a decomposition-based method for the problem using the submodularity of the spread function.
The aforementioned studies consider a single decision maker that chooses a seed set to start a diffusion process and assume there is no competition.

In many real world problems, it is possible to observe competing diffusion processes such as the spread of new technologies of two competing companies, the spread of rumors etc. One class of competitive IMPs integrates multiple seed types in the well-known diffusion models \citep{carnes2007maximizing,borodin2010threshold}. In such problems, multiple kinds of information spread over the network simultaneously. The objective is to maximize the spread of a specific type by choosing an optimal seed given the seeds of other types (i.e., to determine an optimal response to the decision of the competitors). In \citet{chen2011influence}, a competing idea emerges during the diffusion as a result of product quality. In another class of problems, the aim is to minimize the opponent's spread or to minimize the size of the seed set \citep{Budak,Nguyen,he2012influence}. In these studies, the decision maker is again the player who is the second mover, i.e., there is an initial seed of the competitor which may be known or unknown to the decision maker before his/her choice.

In this study, we consider a competitive environment in the form of a Stackelberg game, which is also called a leader-follower game. There are only a few related works in the literature in which the problem is studied from the perspectives of two players. All of them consider two competing players who want to maximize their spreads. \cite{bharathi2007competitive} provide an approximation algorithm for the second player under the Independent Cascade model and briefly discuss a strategy of the first player. \cite{kostka2008word} analyze the same problem under a particular deterministic diffusion model by making use of location theory concepts. They show that finding even an approximate solution for the first player is $\mathcal{NP}$-complete. \cite{clark2011maximizing} assume a Markov diffusion model and show that both players' problems have submodular objective function under this model. Unlike the previous studies, \citet{Hemmati}, consider a bilevel model in which both players have the same objective function but with opposite directions. The first player seeks to minimize the spread by protecting some nodes whereas the second player wants to maximize it by targeting unprotected nodes. The authors adopt a deterministic linear threshold diffusion process.

The most relevant work to ours is due to \citet{Hemmati}. Our bilevel model differs in the diffusion process: we employ a stochastic linear threshold model where nodes have uncertain thresholds as defined in \citet{Kempe}. The first player selects a subset of nodes referred to as \emph{seed nodes} to maximize the spread, whereas the second player deactivates some of the seed nodes to minimize the same measure. An example of such a diffusion process is the misinformation spread by an antagonist or adversary organization with the purpose of causing chaos in the society. In this context, solving the follower's problem answers the question of which individuals among a group of initial spreaders, i.e., the seed, must be prevented or restrained from influencing other people. Solving the problem of the leader, on the other hand, corresponds to finding the strategy that maximizes the spread in the existence of a rational follower. In other words, the seed selected by the leader identifies the individuals that are most likely to be targeted by the antagonist. Hence, solving the problem of the leader helps to pinpoint the key people in the social network. To this end, we formulate the problem as a bilevel programming model, and propose methods for its solution. To the best of our knowledge, our study is the first one that focuses on a Stackelberg game between two players for influence maximization problem under a stochastic linear threshold diffusion model.


The remainder of the paper is organized as follows. We develop a stochastic discrete bilevel programming formulation of the problem in Section \ref{section:Problem Definition}. In Section \ref{section:Solution Methods}, we provide an approximation scheme for the follower's problem using the Sample Average Approximation (SAA) method \citep{mak1999monte,Norkin1998}, and then we employ it as a subroutine within a metaheuristic. This gives rise to a  matheuristic for the solution of the overall problem. Section \ref{section:Numerical Results} includes the computational results that are based on the experimental setting explained in the same section. Some future research directions are suggested in the last section.

\section{Problem Definition}\label{section:Problem Definition}

IMP is defined as finding an initial set of $k$ nodes to start a diffusion process in a social network so that the total number of affected nodes at the end of the process is maximized. The problem considered in this study is a competitive version of the IMP, which can be regarded as a Stackelberg game between two players. Given a directed graph (or digraph) $D=(V,A)$, the first player (leader) determines a subset $S \subset V$ of nodes (individuals) to activate in order to propagate a belief, idea, or campaign. The activated nodes, also called the seed set, are not only influenced directly as a result of the campaign, but they also have the capability of influencing other individuals. Given that the leader decides on the seed set, the second player (the follower) having perfect information on $S$ \textit{deactivates} some nodes $P \subset S$ so that they cannot influence other nodes, which has the consequence of keeping the number of activated nodes at a low level. Deactivation can be achieved in different ways depending on the problem context. The follower may convince a seed node not to spread the idea by giving promotions, making a payment and so on, or can also remove its links to other nodes. After the follower's decision, the nodes that are activated by the leader and not deactivated by the follower ($S \setminus P$), influence other nodes in the network according to the well known Linear Threshold (LT) diffusion model, where a node becomes influenced only if the total weight on the incoming arcs from its influenced neighbors exceeds a random threshold value. The goal of the leader is to maximize the expected number of influenced nodes whereas the follower tries to minimize it. We refer to the problem as the Influence Maximization Problem with Deactivation (IMPD).

\begin{figure}[H]
\centering
  \begin{subfigure}[b]{0.50\linewidth}
  \centering
  \fbox{
    \begin{tikzpicture}[->,>=stealth',shorten >=1pt,auto,node distance=2.5cm,
                    thick, main node/.style={circle,draw,font=\sffamily\Large\bfseries}]
\tikzstyle{every state}=[fill=white,draw=black,text=black,scale=0.9]

  \node[state,fill=black!20!white] 		   (A)             		  {A,0.3};
  \node[state]         (B) [right of=A] 	  {B,0.6};
  \node[state]         (C) [right of=B]       {C,0.8};
  \node[state,fill=black!20!white]         (D) [below of=A] {D,0.4};
  \node[state]         (E) [right of=D] {E,0.1};
  \node[state]         (F) [right of=E] {F,0.5};
  \path (A) edge			  node {0.7} (B)
        (B) edge 			  node {0.1} (C)
        	edge			  node[above, yshift=5pt] {0.1} (D)
        	edge			  node[left] {0.2} (E)
        (C) edge              node {0.6} (F)
        (D) edge              node {0.1} (A)
        (E) edge 			  node[above, yshift=5pt] {0.9} (C)
    	    edge			  node[below] {0.2} (D)
    	    edge			  node[below] {0.2} (F)
        ;
\end{tikzpicture}
}
\caption{IMP optimal seed} \label{fig:a}
\end{subfigure}%
  \begin{subfigure}[b]{0.50\linewidth}
  \centering  \fbox{
    \begin{tikzpicture}[->,>=stealth',shorten >=1pt,auto,node distance=2.5cm,
                    thick, main node/.style={circle,draw,font=\sffamily\Large\bfseries}]
\tikzstyle{every state}=[fill=white,draw=black,text=black,scale=0.9]

  \node[state,dashed] 		   (A)             		  {A,0.3};
  \node[state]         (B) [right of=A] 	  {B,0.6};
  \node[state]         (C) [right of=B]       {C,0.8};
  \node[state,fill=black!20!white]         (D) [below of=A] {D,0.4};
  \node[state]         (E) [right of=D] {E,0.1};
  \node[state]         (F) [right of=E] {F,0.5};
  \path (A) edge			  node {0.7} (B)
        (B) edge 			  node {0.1} (C)
        	edge			  node[above, yshift=5pt] {0.1} (D)
        	edge			  node[left] {0.2} (E)
        (C) edge              node {0.6} (F)
        (D) edge              node {0.1} (A)
        (E) edge 			  node[above, yshift=5pt] {0.9} (C)
    	    edge			  node[below] {0.2} (D)
    	    edge			  node[below] {0.2} (F)
        ;
\end{tikzpicture}
}
\caption{Spread of IMP optimal seed after deactivation} \label{fig:b}
\end{subfigure}
\begin{subfigure}[b]{0.50\linewidth}
  \centering  \fbox{
 \begin{tikzpicture}[->,>=stealth',shorten >=1pt,auto,node distance=2.5cm,
                    thick, main node/.style={circle,draw,font=\sffamily\Large\bfseries}]
\tikzstyle{every state}=[fill=white,draw=black,text=black,scale=0.9]

  \node[state] 		   (A)             		  {A,0.3};
  \node[state,fill=black!20!white]         (B) [right of=A] 	  {B,0.6};
  \node[state,fill=black!20!white]         (C) [right of=B]       {C,0.8};
  \node[state]         (D) [below of=A] {D,0.4};
  \node[state]         (E) [right of=D] {E,0.1};
  \node[state]         (F) [right of=E] {F,0.5};
  \path (A) edge			  node {0.7} (B)
        (B) edge 			  node {0.1} (C)
        	edge			  node[above, yshift=5pt] {0.1} (D)
        	edge			  node[left] {0.2} (E)
        (C) edge              node {0.6} (F)
        (D) edge              node {0.1} (A)
        (E) edge 			  node[above, yshift=5pt] {0.9} (C)
    	    edge			  node[below] {0.2} (D)
    	    edge			  node[below] {0.2} (F)
        ;
\end{tikzpicture}
}
\caption{IMPD optimal seed} \label{fig:c}
\end{subfigure}%
\begin{subfigure}[b]{0.50\linewidth}
  \centering  \fbox{
 \begin{tikzpicture}[->,>=stealth',shorten >=1pt,auto,node distance=2.5cm,
                    thick, main node/.style={circle,draw,font=\sffamily\Large\bfseries}]
\tikzstyle{every state}=[fill=white,draw=black,text=black,scale=0.9]

  \node[state] 		   (A)             		  {A,0.3};
  \node[state,dashed]         (B) [right of=A] 	  {B,0.6};
  \node[state,fill=black!20!white]         (C) [right of=B]       {C,0.8};
  \node[state]         (D) [below of=A] {D,0.4};
  \node[state]         (E) [right of=D] {E,0.1};
  \node[state,fill=black!20!white]         (F) [right of=E] {F,0.5};
  \path (A) edge			  node {0.7} (B)
        (B) edge 			  node {0.1} (C)
        	edge			  node[above, yshift=5pt] {0.1} (D)
        	edge			  node[left] {0.2} (E)
        (C) edge              node {0.6} (F)
        (D) edge              node {0.1} (A)
        (E) edge 			  node[above, yshift=5pt] {0.9} (C)
    	    edge			  node[below] {0.2} (D)
    	    edge			  node[below] {0.2} (F)
        ;
\end{tikzpicture}
}
\caption{Spread of IMPD optimal seed after deactivation} \label{fig:d}
\end{subfigure}
\caption{Comparison of two leader solutions under deterministic thresholds assumption}
\label{Figure:IMPD example}
\end{figure}
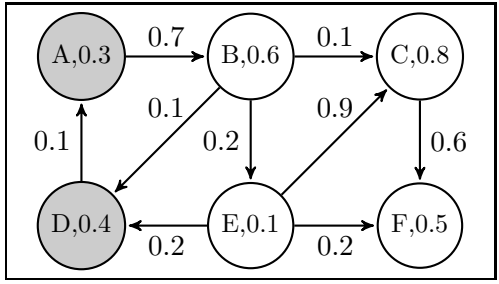
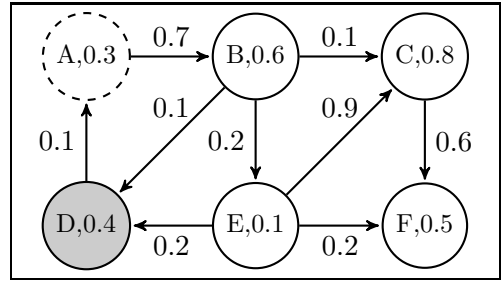
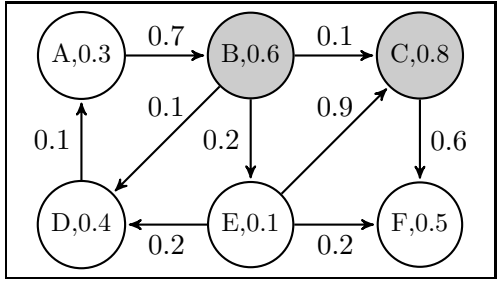
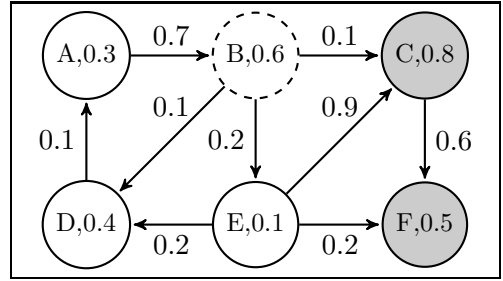

An illustration of the problem is presented in Figure \ref{Figure:IMPD example} for two different leader decisions under the assumption that node thresholds are deterministic. Threshold values are shown next to the node names and arc weights are shown on the arcs. In Figure \ref{fig:a} and Figure \ref{fig:c} the leader activates the optimal seed nodes of IMP and IMPD, respectively, which are displayed by shaded circles. (The resulting number of influenced nodes would be 6 and 4 if there were no deactivation.) Then the follower deactivates one of the seed nodes. The deactivated nodes are displayed as dashed circles and the influenced nodes at the end of diffusion process are shaded in Figures \ref{fig:b} and \ref{fig:d}. As can be observed, the IMP optimal seed cannot preserve its superiority in the existence of a rational opponent.

Stackelberg games can be formulated as bilevel programs where the problem of the follower is a constraint set in the leader's problem. On the other hand, the uncertainty of node thresholds in the LT diffusion model turns our problem into a stochastic optimization problem. Therefore, a discrete stochastic bilevel programming model for the IMPD is proposed in \eqref{equation:expectedSpread}--\eqref{equation:reactionSet}. The sets, parameters, and decision variables used in the formulation are provided below. Note that the model takes into account the possibility that not all individuals (nodes) are alike in terms of activation and deactivation. Therefore, it is assumed that there is a budget for both the leader and the follower, and there is a cost for activation and deactivation associated with each individual. This cost may be the same for each individual giving rise to a cardinality-based IMPD, or it may change with respect to individuals resulting in cost-based IMPD.

\begin{table}[H]
\begin{tabular}{lllp{12cm}}
\multicolumn{4}{l}{{\textit{Sets and Parameters:}}} \\
&$V$	&:& set of all nodes in the network where $|V|=n$ \\
&$c_i$	&:& leader's cost of activating node $i$  \\
&$e_i$	&:& follower's cost of deactivating node $i$ \\
&$\theta_i$	&:& influence threshold of node $i$\\
&$C$ &:& leader's budget to activate nodes \\
&$E$ &:& follower's budget to deactivate nodes\\
\multicolumn{4}{l}{\textit{Decision variables:}}\\
&$x_{i}$	&:& 1 if node $i$ is activated by the leader; 0 otherwise\\
&$y_{i}$	&:& 1 if node $i$ is deactivated by the follower; 0 otherwise
\end{tabular}
\end{table}

Using this notation, the stochastic bilevel programming model for the IMPD can be defined as follows:
\begin{align}
\text{IMPD:}\notag \\
& \max_{\mathbf{x}} \hspace{5pt} \mathbb{E}_{\bm{\theta}} [g(\mathbf{x},\mathbf{y}^{\star},\bm{\theta)}] \label{equation:expectedSpread}\\
&\text{ s.t.} \notag\\
& \indent  \mathbf{c}^T \mathbf{x} \leq C \label{equation:leaderBudget}\\
& \indent \mathbf{x} \in \{0,1\}^n \label{equation:SBP-binaryX}\\
& \indent \mathbf{y}^{\star} \in \arg \min_{\mathbf{y}} \Big \{ \mathbb{E}_{\bm{\theta}} [g(\mathbf{x},\mathbf{y},\bm{\theta)}]:
\mathbf{e}^T \mathbf{y} \leq E, \mathbf{y}\leq \mathbf{x}, \mathbf{y} \in \{0,1\}^n \Big \} \label{equation:reactionSet}
\end{align}
where $g(\mathbf{x},\mathbf{y}^{\star},\bm{\theta)}$ represents the number of influenced nodes for a given threshold realization (scenario) vector $\bm{\theta}$ corresponding to a given seed selection strategy $\mathbf{x}$ of the leader and optimal seed deactivation strategy $\mathbf{y}^{\star}$ of the follower. For a given realization, vector $\bm{\theta}$ is known, which enables the computation of $g(\mathbf{x},\mathbf{y}^{\star},\bm{\theta)}$ in an algorithmic way bounded by polynomial time using the LT diffusion model. $\mathbb{E}_{\bm{\theta}} [g(\mathbf{x},\mathbf{y}^{\star},\bm{\theta)}]$ denotes the spread, i.e., the expected number of influenced nodes where the expectation is taken over the probability distribution of the threshold vector $\bm{\theta}$ the components of which are the threshold values $\theta_{i}$ of the nodes. Note that the optimization problem of the follower given by \eqref{equation:reactionSet} becomes a constraint in the leader's problem. In other words, the leader tries to give the best decision $\mathbf{x}$ with the anticipation that the follower gives the best response $\mathbf{y}^{\star}$.

The objective function \eqref{equation:expectedSpread} of the leader is to maximize the spread. Constraint \eqref{equation:leaderBudget} is the budget constraint of the leader. Constraint \eqref{equation:SBP-binaryX} puts binary restriction on variables $\mathbf{x}$. Constraint set \eqref{equation:reactionSet} states that $\mathbf{y}^{\star}$ must be in the rational reaction set (optimal response) of the follower. The first constraint of this set ($\mathbf{e}^T \mathbf{y} \leq E$) limits the deactivation cost by the available budget. The second one, $\mathbf{y} \leq \mathbf{x}$, ensures that a node cannot be deactivated unless it is activated by the leader. The last one is the binary restriction on variables $\mathbf{y}$. The rational reaction set includes $\mathbf{y}$'s that satisfy the aforementioned three constraints and minimize the spread for a given $\mathbf{x}$.

There is no closed-form expression to calculate $\mathbb{E}_{\bm{\theta}} [g(\mathbf{x},\mathbf{y}^{\star},\bm{\theta)}]$. However, it is possible to develop a deterministic formulation equivalent to \eqref{equation:expectedSpread}--\eqref{equation:reactionSet} that is obtained by enumerating each possible realization of the uncertain threshold vector, under the assumption that the number of threshold realizations is finite. The additional notation needed for the so-called deterministic equivalent formulation (DEF) is given below.
\vspace{10pt}

\begin{tabular}{lllp{12cm}}
\multicolumn{4}{l}{{\textit{Sets and Parameters:}}} \\
&$R$	&:& set of all possible threshold realizations \\
&$\theta_{ir}$	&:&  threshold of node $i$ in threshold realization $r$\\
&$w_{ij}$	&:& weight of link $(i,j)$, 0 if the link does not exist\\
&$p_r$		&:& probability of threshold realization $r$ \\
\multicolumn{4}{l}{{\textit{Decision variable:}}}\\
&$u_{ir}$	&:& 1 if node $i$ is influenced in threshold realization $r$; 0 otherwise
\end{tabular}
\vspace{0.8cm}

As a result, IMPD-DEF can be written as follows:
\begin{align}
\text{IMPD-DEF:}\notag \\
&z^{*}_L=\displaystyle\max_{\textbf{x}} z(\textbf{x}) \label{equation:zD}\\
&\indent\text{s.t.} \notag\\
& \indent  \displaystyle\sum_{i \in V} c_{i} x_{i} \leq C \label{equation:budgetD}\\
& \indent x_{i}\in \{0,1\} \indent &i& \in V \label{equation:binaryX}
\end{align}
where 
\begin{align}
&  z(\textbf{x}) = \min_{\textbf{y},\textbf{u}} \sum_{r \in R}\sum_{i \in V}p_r u_{ir}  \label{equation:zA}\\
& \indent \text{\hspace{15pt}s.t.} \notag\\
& \indent\indent \displaystyle\sum_{i \in V} e_{i} y_{i} \leq E \label{equation:budgetA}\\
& \indent\indent y_{i}\leq x_{i}  &i& \in V \label{equation:relateXY}\\
& \indent\indent u_{ir} \geq  x_i-y_i  &i& \in V, r \in R \label{equation:seedSet}\\
& \indent\indent u_{ir}+y_i \geq \displaystyle\sum_{j \in V}w_{ji}u_{jr}-\theta_{ir} + \epsilon &i& \in V, r \in R \label{equation:threshold}\\
& \indent\indent u_{ir},y_{i} \in \{0,1\} &i& \in V, r \in R\label{equation:binaryA}
\end{align}
Here, \eqref{equation:zD}--\eqref{equation:binaryX} represents the leader's upper level problem (ULP), and \eqref{equation:zA}--\eqref{equation:binaryA} is the follower's lower level problem (LLP). The objective function $z(\mathbf{x})=\min_{\textbf{y},\textbf{u}} \sum_{r \in R}\sum_{i \in V}p_r u_{ir}$ denotes the minimum average number of influenced nodes for a given leader decision $\mathbf{x}$ and optimal follower decision $\mathbf{y}$ corresponding to $\mathbf{x}$. Note that it is an approximation to the value of the spread. Inequality \eqref{equation:budgetD} is the budget constraint of the leader. Decision variables $x_i$ take binary values by constraints \eqref{equation:binaryX}. Constraint \eqref{equation:budgetA} in the LLP is the budget constraint of the follower, and constraints \eqref{equation:relateXY} imply that a node can only be deactivated by the follower if it is activated by the leader. Constraints \eqref{equation:seedSet} force the seed nodes that are not deactivated $(x_i=1, y_i=0)$ to be influenced by default, i.e., $u_{ir}=1$. In other cases, i.e., $(x_i=0, y_i=0)$ and $(x_i=1, y_i=1)$, the value of $u_{ir}$ is determined by constraint \eqref{equation:threshold} and the objective function. This is achieved as follows: The summation on the right-hand side (RHS) of constraints \eqref{equation:threshold} is the total weight on the incoming links to node $i$ from its influenced predecessors. This total weight cannot exceed one by the definition of link weights in the LT diffusion model. If the total weight is equal to or exceeds the threshold $\theta_{ir}$ of node $i$ in realization $r$, the RHS becomes positive between zero and one, which forces the left-hand side to be at least one. In that case, if $i$ is a deactivated node $(x_i=1,y_i=1)$, $u_{ir}$ is set to zero due to the minimization objective (satisfying the assumption that a deactivated node cannot be influenced). Otherwise, $u_{ir}$ is set to one because $(x_i=0,y_i=0)$, i.e., node $i$, which is not a seed node, is influenced in realization $r$. If the RHS is non-positive, then a node is not influenced due to the minimization of the objective function. $\epsilon$ is a small positive number and guarantees that the RHS is positive when the total weight is equal to or greater than  the threshold. Lastly, $u_{ir}$ and $y_i$ take binary values by constraints \eqref{equation:binaryA}. Note that, if the values $\theta_{ir}$ of the threshold parameter were deterministic, then the realization index $r$ would disappear from the model \eqref{equation:zA}--\eqref{equation:binaryA} and it would be a much simpler problem to solve.

\section{Solution Method}\label{section:Solution Methods}
Bilevel problems are known to be $\mathcal{NP}$-hard \citep{bard1991some, ben1990computational}. In case of a linear LLP with continuous decision variables, a bilevel problem can be reformulated as a single level problem using the Karush-Kuhn-Tucker (KKT) optimality conditions \citep{colson2005bilevel, lu2016multilevel}. If the objective functions of the LLP and ULP are the same, it is also possible to write the dual formulation of the LLP to obtain a single-level formulation. In our problem, both the ULP and LLP include binary decision variables, which prevents the aforementioned approaches. One method to solve discrete bilevel optimization problems is to develop a heuristic method for the leader's ULP, which requires to solve the follower's LLP \eqref{equation:zA}--\eqref{equation:binaryA} to optimality for each candidate solution of the leader that is generated. We propose two matheuristics based on an approximation of the follower's problem that is explained below.
\subsection{Sample Average Approximation Method for the Follower's Lower Level Problem}
In the LT diffusion model, each of the threshold parameter values, $\theta_{i}$, follows a continuous uniform distribution \citep{Kempe}. Therefore, it is not possible to enumerate all possible realizations. Even if $\theta_{i}$'s had a discrete distribution, it would require the solution of a large-sized integer problem in the lower level, leading to an inefficient solution approach. Sample Average Approximation (SAA) method can be utilized in such a case by using random samples that are generated from the set of all realizations. We would like to mention that our application of the SAA method is based on the framework proposed by \citet{kleywegt2002sample}. This SAA implementation consists of three stages. The first stage employs the solution of the follower's approximate LLP, referred to as ALLP, which is formulated below. In ALLP, $N$ represents the sample size, namely the number of threshold realizations. Notice that each realization $r$ consists of $|V|$ threshold values, $\theta_{i}$, associated with each node $i \in V$.
\begin{align}
&  \mathrm{ALLP:} \hat{z}_{N}(\mathbf{x}) = \displaystyle\min_{\textbf{y},\textbf{u}} \frac{1}{N}\displaystyle\sum_{r=1}^N \sum_{i \in V}u_{ir}  \label{equation:SAA-zA}\\
& \indent \text{\hspace{15pt}s.t.} \notag\\
& \indent\indent \eqref{equation:budgetA},\eqref{equation:relateXY}, \mathrm{and} \notag \\
& \indent\indent u_{ir} \geq  x_i-y_i  &i& \in V, r=1,...,N \label{equation:SAA-threshold1}\\
& \indent\indent u_{ir}+y_i \geq \displaystyle\sum_{j \in V}w_{ji}u_{jr}-\theta_{ir} + \epsilon &i& \in V, r=1,...,N \label{equation:SAA-threshold2}\\
& \indent\indent u_{ir},y_{i} \in \{0,1\} &i& \in V, r=1,...,N \label{equation:SAA-binaryA}
\end{align}
The optimal objective value $\hat{z}_{N}(\mathbf{x})$ of the ALLP for a given leader decision $\mathbf{x}$ is a negatively biased estimator of the optimal objective value $z(\mathbf{x})$ of the original LLP. Namely, $\mathbb{E}[\hat{z}_{N}(\mathbf{x})] \leq z(\mathbf{x})$ \citep{mak1999monte,Norkin1998}. In order to approximate the objective value $z(\mathbf{x})$ from above, we generate $M$ independently and identically distributed threshold samples of size $N$, and solve the ALLP for each sample $m$ (also called batch $m$) to obtain the optimal follower decision $\tilde{\mathbf{y}}^m$ with the objective value $\hat{z}_{N}^m (\mathbf{x})$. The average of these objective values computed in expression \eqref{Equation:Z_NM} constitutes a statistical lower bound on the optimal spread $z(\mathbf{x})$ because of the negative bias of each $\hat{z}_{N}^{m}(\mathbf{x})$ as discussed before.
\begin{equation}
\bar{z}_N^M (\mathbf{x})=\frac{1}{M}\sum_{m=1}^M \hat{z}_{N}^m (\mathbf{x}).
\label{Equation:Z_NM}
\end{equation}
This completes the first stage of the SAA method. Notice that the decision variables $\mathbf{y}$ representing the follower decision in the ALLP are not dependent on threshold realizations, and hence do not have index $r$. Therefore, they represent the first-stage variables in the context of two-stage stochastic programming models for which the SAA method is developed.

The second stage of the SAA method involves determining the best follower response $\tilde{\mathbf{y}}^*$ among $\tilde{\mathbf{y}}^m, m \in M$ that were found in the first stage. To this end, the ALLP can be solved for each batch $m$ by fixing the optimal follower decision $\tilde{\mathbf{y}}^m$ associated with that batch and using a larger number of threshold realizations $N' \gg N$. Notice that the only decision variables remaining in the ALLP model are the second-stage variables $u_{ir}$ since leader's decision $\mathbf{x}$ is also given in the LLP. However, we apply a more efficient procedure and calculate the resulting objective value of each batch $m$, denoted as $\hat{z}_{N'}(\mathbf{x},\tilde{\mathbf{y}}^m)$, in an iterative way by fixing the activated nodes $\mathbf{x}$ and deactivated nodes $\tilde{\mathbf{y}}^m$ in the network, and simulating the influence propagation until no more nodes are influenced. Our experiments revealed that this numerical simulation of computing the spread is faster than solving the mathematical model. The best follower decision $\tilde{\mathbf{y}}^*$ for a given leader decision $\mathbf{x}$ is found as the one $\tilde{\mathbf{y}}^m$ that provides the smallest spread value $\hat{z}_{N'}(\mathbf{x},\tilde{\mathbf{y}}^m)$, namely $\hat{z}_{N'}(\mathbf{x},\tilde{\mathbf{y}}^*)=\min_{m} \hat{z}_{N'}(\mathbf{x},\tilde{\mathbf{y}}^m)$. \citet{kleywegt2002sample} show that $\hat{z}_{N'}(\mathbf{x},\tilde{\mathbf{y}}^{\star})$ converges to $z(\mathbf{x})$ with probability one as $N' \rightarrow \infty$.

The third and last stage of the SAA method consists of computing an unbiased estimate of the optimal spread $z(\mathbf{x})$. This is achieved by taking an independent sample size $N'' \gg N'$ \citep{verweij2003sample,Guney2017}. Note that this again requires either the solution of the ALLP in terms of second-stage variables $u_{ir}$ by fixing leader's decision variable $\mathbf{x}$ and follower's first-stage decision variable $\tilde{\mathbf{y}}^*$ or adopting the numerical simulation. As before, the latter approach turns out to be more efficient. It is clear that $\hat{z}_{N''}(\mathbf{x},\tilde{\mathbf{y}}^*) \geq z(\mathbf{x})$, as the former objective value is associated with a feasible solution, while the latter is the optimal objective value of the LLP. In other words, we obtain an upper bound on the optimal objective value $z(\mathbf{x})$ of the LLP in IMPD-DEF.

In the SAA method, it is possible to calculate an estimator of the gap for the optimal objective value of the stochastic integer programming model in question. In our case, the optimality gap for $z(\mathbf{x})$ can be estimated by the quantity $\hat{z}_{N''}(\mathbf{x},\tilde{\mathbf{y}}^*) - \bar{z}_N^M(\mathbf{x})$. Besides the estimation of this gap, we can also compute the variance of the estimated gap as follows:
\begin{equation}
\begin{split}
V \big( \hat{z}_{N''}(\mathbf{x},\tilde{\mathbf{y}}^*) - \bar{z}_N^M(\mathbf{x}) \big ) & =V \big ( \hat{z}_{N''}(\mathbf{x},\tilde{\mathbf{y}}^*) \big ) + V \big ( \bar{z}_N^M(\mathbf{x}) \big ) \\
& =\frac{1}{N''(N''-1)}\sum_{r=1}^{N''} \big[ \hat{z}_{N''}(\mathbf{x},\tilde{\mathbf{y}}^*)-  g(\mathbf{x},\mathbf{y}^*,\theta_r) \big] ^2 + \\
& \frac{1}{M(M-1)}\sum_{m=1}^{M} \big[ \hat{z}_{N}^m (\mathbf{x})-\bar{z}_N^M (\mathbf{x}) \big]^2
\end{split}
\label{gapvar}
\end{equation}
We provide in Algorithm \ref{SAA} the steps of the SAA method implemented for the approximate solution of the LLP in IMPD-DEF for a given leader strategy $\mathbf{x}$. Lines 1--6 correspond to stage 1 of the algorithm, where lower bound $\bar{z}_N^M$ and candidate solutions $\tilde{\mathbf{y}}^m$ are generated.  Lines 7--11 comprise stage 2 where the best response $\tilde{\mathbf{y}}^*$ is determined. The remaining lines form the third stage, which involves the computation of the upper bound $\hat{z}_{N''}(\mathbf{x},\tilde{\mathbf{y}}^*)$ as well as an estimate of the optimality gap, and its variance. For solving the ULP of the leader, we devise three approaches in the following subsections.
\begin{algorithm}[h!]
\caption{The approximate computation of the LLP for a given leader strategy $\mathbf{x}$} \label{SAA}
\begin{algorithmic}[1]
\STATE {Choose sample sizes $N, N', N''$, and the number of batches $M$}
\FOR{$m=1,\dots , M $}
\STATE {Generate $N$ threshold realizations}
\STATE {Solve ALLP to obtain the optimal objective value $\hat{z}_{N}^m (\mathbf{x})$ and optimal solution $\tilde{\mathbf{y}}^m$}
\ENDFOR
\STATE {Compute $\bar{z}_N^M (\mathbf{x})=\frac{1}{M}\sum_{m=1}^M \hat{z}_{N}^m (\mathbf{x})$}
\STATE {Generate $N'$ threshold realizations}
\FOR{$m=1,\ldots , M $}
\STATE {Compute $\hat{z}_{N'}(\mathbf{x},\tilde{\mathbf{y}}^m)$ by numeric simulation }
\ENDFOR
\STATE {Choose the best solution $\tilde{\mathbf{y}}^*=\arg \min_{m} \hat{z}_{N'}(\mathbf{x},\tilde{\mathbf{y}}^m)$}
\STATE {Generate $N''$ threshold realizations}
\STATE {Compute $\hat{z}_{N''}(\mathbf{x},\tilde{\mathbf{y}}^*)$ by numeric simulation}
\STATE {Compute the optimality gap $\hat{z}_{N''}(\mathbf{x},\tilde{\mathbf{y}}^*) - \bar{z}_N^M(\mathbf{x})$ and its variance given in \eqref{gapvar}}
\end{algorithmic}
\end{algorithm}

\subsection{Complete Enumeration of the Leader's Solutions}
Note that we solve the LLP in IMPD-DEF given in \eqref{equation:zA}--\eqref{equation:binaryA}, which is a stochastic integer program, in an approximate way using the SAA method. Since the estimated objective value of the follower is given as $\hat{z}_{N''}(\mathbf{x},\tilde{\mathbf{y}}^{\star})$ for a given leader strategy $\mathbf{x}$, we can rewrite the ULP as follows:
\begin{equation}
\hat{z}_L=\max_{\mathbf{x} \in \mathbb{X}} \hat{z}_{N''}(\mathbf{x},\tilde{\mathbf{y}}^*) \label{ULP}
\end{equation}
where $\mathbb{X}=\{ \mathbf{x}: \mathbf{c}^T \mathbf{x} \leq C, \mathbf{x} \in \{0,1\} \}$.

One method to solve \eqref{ULP} to optimality is to enumerate all feasible solutions of the leader's problem ULP. Then, the estimated optimal objective value of the follower, $\hat{z}_{N''}(\mathbf{x},\tilde{\mathbf{y}}^*)$, is computed using Algorithm \ref{SAA} for each feasible solution. For the sake of notation, we will denote this value as $\hat{z}_{SAA}(\mathbf{x})$ in the sequel. Note that it is nondecreasing in the number of activated nodes, i.e., the seed size, which allows the elimination of the solutions that are not maximal. Let $\mathbb{X'}$ denote the set of all maximal elements of $\mathbb{X}$ ($A \in \mathds{X}$ is maximal if there is no $B\in \mathbb{X}$ such that $A \subset B$). Hence, we can consider to solve the problem $\hat{z}_L=\max_{\mathbf{x} \in \mathbb{X}'} \hat{z}_{SAA}(\mathbf{x})$. Obviously, the activation budget $C$ and the network size $n$ determine the size of $\mathbb{X}'$. The number of candidate feasible solutions to be evaluated increases exponentially with increasing values of these parameters. Therefore, this method can only be helpful for very small networks. In this study, the results of the complete enumeration method are used to evaluate the performance of the matheuristic methods for small instances.

\subsection{Matheuristic Methods}\label{subsection:Matheuristic Methods}
We propose two matheuristics based on Simulated Annealing and Tabu Search metaheuristics. Both of these matheuristics perform a search in the solution space of the leader's decision variables $\mathbf{x}$ and for each solution visited, the optimal spread is estimated using Algorithm \ref{SAA}. Therefore, each method can be regarded as a metaheuristic coupled with the solution of a stochastic mathematical programming model solved by means of SAA. There are two important aspects with regard to these methods. First, only feasible solutions are considered due to the costly $\hat{z}_{SAA}(\mathbf{x})$ computation of the LLP. Moreover, it is not allowed to visit the same solution more than once for the sake of computational efficiency. This is achieved by using an explicit memory structure that stores each solution generated by means of a hash function.

Two approaches are considered for initial solution generation. The first one is based on the influence spread caused by each node as well as the activation and deactivation costs of each node. The spread, also called the score of node $i$, is calculated by solving the ALLP with $N$ threshold realizations in which $x_i=1$, $x_j=0$  for $j \neq i$, and $y_i=0$. This implies that node $i$ is the only active node and it is not deactivated by the follower. After calculating the scores for all nodes, they are sorted in nonincreasing  order. If it is a cost-based IMPD instance, then the second half of the nodes are removed from the list, while the first half is sorted again in terms of increasing value of the $c_{i}/e_{i}$ ratio. The nodes at the top of the final list are selected as seed nodes until the budget constraint is violated. This selection favors relatively more influential nodes that can be activated at a low cost but deactivated at a high cost. The second approach for initial solution generation is based only on the activation and deactivation costs of the nodes. In this approach, nodes are sorted in nondecreasing order with respect to their activation costs $c_i$. Ties are broken in favor of the largest deactivation cost $e_i$.

Before going into the details of the matheuristics, we would like to point out that the objective value of a candidate leader solution $S=\{i \in V: x_i=1\}$ is computed as
\begin{equation}
f(S)=\hat{z}_{SAA}(S)+  \frac{C-\sum_{i \in S} c_i }{\bar{c}} , \label{objfnc}
\end{equation}
where $\bar{c}$ is the average activation cost of the nodes in the network. The second term on the right-hand side of Equation \eqref{objfnc} represents the average number of nodes that can be activated using the remaining budget for a solution $S$. The rationale of using this pseudo objective value is to reward a seed set that achieves a lower spread in comparison with another seed set albeit at a lower activation cost. This implies that we also take into account the potential of a seed set to improve the spread. Note that this approach does not favor an unpromising solution under the assumption that for any feasible solution there exist nodes in the network not belonging to the seed set and having an activation cost less than $\bar{c}$. This is not an unreasonable assumption since the size of the seed set is much smaller than the number of nodes in the network.

\subsubsection{Simulated Annealing Based Matheuristic} \label{Section:SA}
The first method we propose is a matheuristic based on Simulated Annealing (SA) referred to as SAM. At each iteration of SAM, an eligible solution (explained in detail below) is selected randomly from the neighborhood of the current solution. The latter is updated if either the neighbor has a better objective value (higher spread) or the neighbor has a worse objective value but still accepted as the current solution with a probability that is a function of the amount of the deterioration in the objective value and the current temperature. Let $\mathbf{x}_{SAM}^{\ast}$ denote the best leader strategy that SAM finds, and $\hat{z}_{SAM}^{\ast}$ denote the corresponding objective value. Then, $\hat{z}_{SAM}^{\ast}$ is a lower bound on $\hat{z}_L$ given in Equation \eqref{ULP}. 

Before moving on to the details of SAM, we give the definition of an eligible solution.  A solution or seed set $S$ is said to be \textit{eligible} if it is feasible ($\sum_{i \in S} c_i \leq C$), not generated before, and has a positive spread. The last property implies that all the nodes of an eligible solution cannot be deactivated by the follower, namely $\sum_{i \in S} e_i > E$.\\
\noindent\textit{Neighborhood Structure:} Three types of move operators are implemented: 1-Add, 1-Drop, and 1-Swap. A 1-Add move randomly chooses a node $i \in V \setminus S$ such that $S'=S \cup i$ is an eligible solution. A 1-Drop move randomly chooses a node $i\in S$ such that $S'=S \setminus i$ is eligible. Finally, a 1-Swap move randomly exchanges two nodes, $i \in S$ and $j\in V \setminus S$ to generate a new eligible solution $S'=(S \setminus i)\cup j$. All three move operators have the same chance of being selected, i.e., are chosen with equal probability 1/3. If there does not exist an eligible solution in the selected move operator, then the remaining ones are tried.\\
\noindent\textit{Initial Temperature and Temperature Update:} The initial temperature is calculated based on the approach proposed in \citet{Ohlmann07}. In this approach, a number of random solutions are generated, and from the neighborhood of each solution, a random neighbor is created. After the absolute value of the difference between each solution and its neighbor is computed, the average absolute difference is computed, which essentially reflects the objective value changes in the landscape of the objective function. The initial temperature to be used is determined by assuming that a random neighbor of the current solution whose objective value is worse than that of the current solution and equal to the average absolute difference is accepted with an initial probability of acceptance, $p_0$. In our implementation we generate 20 solutions. At the end of each cycle, we check whether the proportion of accepted solutions is greater than a threshold value $\phi$. If this is the case, then the current temperature is reduced to half, as the number of accepted solutions turns out to be large. Otherwise, the temperature is updated in a geometric fashion using a cooling ratio of $r$, as is done frequently in the literature. In other words, $T \leftarrow rT$.\\
\noindent\textit{Cycle Length:} The number of iterations in each cycle, i.e., the cycle length $L$ is dynamic. The initial cycle length $L_0$ is the average neighborhood size of the initial solution obtained by the two approaches explained earlier, over the three move types (e.g., $L_0=(k+(n-k)+(n-k)k)/3$ for an initial seed of size $k$). As the temperature decreases throughout the iterations, the acceptance probability of worsening solutions decreases, which makes finding and accepting better solutions more difficult. Therefore, the number of iterations is increased gradually by setting $L \leftarrow (1+\gamma ) L$, where $\gamma \in (0,1)$. \\
\noindent\textit{Termination Criterion:} TSM is executed for a time limit equal to $t_{\max}$. However, if there is no eligible neighbor in any type of moves at an iteration, then the algorithm is stopped. This is more likely in small instances for which the neighborhood sizes are small and at low temperatures since the probability of acceptance is small.

The pseudocode of SAM is displayed as Algorithm \ref{SA}.

\begin{algorithm}[t]
\caption {SA-based Matheuristic (SAM)}\label{SA}
\begin{algorithmic}[1]
\STATE {Compute initial temperature $T$}
\STATE {Generate the initial solution (seed set) $S$ and set $S_{SAM}^{\ast} \leftarrow S$, $\hat{z}_{SAM}^{\ast} \leftarrow \hat{z}_{SAA}(S)$}
\STATE {$j \leftarrow 0$}
\STATE {Set the values of $L\leftarrow L_0$, $r=0.9$, $\phi$=0.5, $\gamma=0.2$}
\WHILE {CPU time $\leq t_{\max}$}
\FOR {$l=1$ to $L$}
\STATE {Set $S' \leftarrow \text{1-Add}(S)$ or $S' \leftarrow \text{1-Drop}(S)$ or $S' \leftarrow \text{1-Swap}(S)$ with the same probability}
\STATE {Compute $\Delta=f(S')-f(S)$}
\IF {$\Delta \geq 0$ or ($\Delta \leq 0$ and $e^{\Delta /T} > U(0,1)$)}
\STATE {Update the current solution $S \leftarrow S', j \leftarrow j+1$}
\ENDIF
\IF {$\hat{z}_{SAA}(S') > \hat{z}_{SAM}^{\ast}$}
\STATE {Update the incumbent solution $S_{SAM}^{\ast} \leftarrow S', \hat{z}_{SAM}^{\ast} \leftarrow \hat{z}_{SAA}(S')$}
\ENDIF
\ENDFOR
\IF {$j/L > \phi$}
\STATE {$T \leftarrow T/2$}
\ELSE
\STATE {$T \leftarrow rT$}
\ENDIF
\STATE {Set $L \leftarrow (1+\gamma)L$, $j \leftarrow 0$}
\ENDWHILE
\STATE {Return $S_{SAM}^{\ast}$ and $\hat{z}_{SAM}^{\ast}$ }
\end{algorithmic}
\end{algorithm}

\subsubsection{Tabu Search Based Matheuristic}
In this method, the solution space of the leader is explored using Tabu Search (TS) that is implemented with a candidate list strategy,  which helps to reduce the computational effort. At each iteration of TS-based matheuristic (TSM), only a promising subset of the neighboring solutions, i.e., those belonging to the candidate list, is considered. The proportion of the size of this subset within the size of the complete neighborhood is controlled by a parameter, denoted as $\tau$. A preprocessing step is carried out before executing the TSM in order to determine the promising candidates at each iteration. This preprocessing step makes use of the length of the shortest path $\lambda_{ij}$ from node $i$ to node $j$ on $D=(V,A)$ where the length of the arc $(i,j)$ is computed as $-\log(w_{ij})$. It is computed between each pair of nodes in the network using Dijkstra's Algorithm \citep{Dijkstra59}. Larger weights on a path between two nodes indicate higher chance of influencing each other under the LT diffusion model. Hence, a small $\lambda_{ij}$ is an indicator of high direct/indirect influence of node $i$ on node $j$. \\
\noindent\textit{Neighborhood Structure:} The move operators 1-Add, 1-Drop and 1-Swap used in SAM are also employed in TSM so that none of the matheuristics is favored over the other. At each iteration, one of the moves is selected with equal probability. As mentioned before, instead of generating all the neighboring solutions using a move operator, we determine a candidate list and take into this list only promising solutions existing in the neighborhood. This approach is based on the idea that if a node has a smaller chance to be influenced by the seed nodes, then adding this node to the seed set has more potential to improve the spread. Let $S$ be the current seed set. The neighborhood size for the 1-Add move is $|V \setminus S|$. To determine a promising subset of neighboring solutions to be included in the candidate list, a score given as $\sum_{i \in S} \lambda_{ij}$ is assigned first to each eligible neighbor $S^\prime = S \cup j$. Then, the candidates are sorted by their score in nonincreasing order. The first $\tau \cdot |V \setminus S|$ elements in the candidate list are evaluated and the best one is chosen. We evaluate all eligible candidates in the 1-Drop move since the size of this neighborhood is relatively small. Its value is equal to $|S|$ at most. The size of the 1-Swap move operator is $|S| \cdot |V \setminus S|$. The score function for this operator is $\sum_{i \in S} \lambda_{ij}$ for any neighbor $S'=(S\setminus k) \cup j$ (i.e., the candidates are evaluated based on only the node that will be added to the seed). The eligible candidates are first sorted in nonincreasing order in terms of their score. Then, top $\tau \cdot |V \setminus S| \cdot |S|$ in the list are taken into consideration, and the best promising neighbor is determined. Recall that this approach helps us to speed up the search process because a costly objective function evaluation needs to be carried out to estimate the spread at the LLP using the SAA method. \\
\noindent\textit{Tabu Structure:} Since the solutions are mapped to integers and stored in a list using a hash function, all solutions visited before are declared tabu in our TS-based matheuristic. \\
\noindent\textit{Diversification Strategy: } A frequency-based long-term memory is utilized in order to penalize the frequently observed nodes in the seed set. To this end, we keep track of the proportion $\pi_{i}$ of the visited solutions that have node $i$ in the seed set. Each candidate solution $S$ is penalized by adding the term $-\mu \sum_{i \in S} \pi(i)$ to its objective value. \\
\noindent\textit{Termination Criterion:} We allocate a time limit equal to $t_{\max}$ as is the case with SAM. TSM terminates before $t_{\max}$ only if no eligible solutions can be obtained by any move operator.

The pseudocode of TSM is displayed as Algorithm \ref{TS}.

\begin{algorithm}[h!]
\caption {TS-based Matheuristic (TSM)}\label{TS}
\begin{algorithmic}[1]
\STATE {Generate the initial solution (seed set) $S$ and set $S_{TSM}^{\ast} \leftarrow S$, $\hat{z}_{TSM}^{\ast} \leftarrow \hat{z}_{SAA}(S)$}
\STATE {Choose $\tau=0.5$, $\mu=1$}
\WHILE {CPU time $\leq t_{\max}$}
\STATE {Set $S' \leftarrow \text{1-Add}(S,\tau)$ or $S' \leftarrow \text{1-Drop}(S)$ or $S' \leftarrow \text{1-Swap}(S,\tau)$ with equal probability}
\IF {$\hat{z}_{SAA}(S') > \hat{z}_{TSM}^{\ast}$}
\STATE {Update the incumbent solution $S_{TSM}^{\ast} \leftarrow S', \hat{z}_{TSM}^{\ast} \leftarrow \hat{z}_{SAA}(S')$}
\ENDIF
\STATE {Update the parameter $\pi_{i},i=1,\ldots n$}
\ENDWHILE
\STATE {Return $S_{TSM}^{\ast}$ and $\hat{z}_{TSM}^{\ast}$ }
\end{algorithmic}
\end{algorithm}


\section{Computational Results} \label{section:Numerical Results}
The matheuristic methods SAM and TSM were coded in the \texttt{C++} language available within Microsoft Visual Studio 2015 environment. The experiments were carried out on a workstation having an Intel$\textregistered$ Xeon$\textregistered$ E5-2687W CPU, 3.10 GHz processor and 64GB RAM. The operating system was Microsoft Windows 7 Professional and the mixed-integer linear programming solver used was CPLEX Optimization Studio 12.7.

\subsection{Verification of the SAA Method} \label{subsection:SAA Method Results}
The performances of SAM and TSM depend heavily on the SAA method because the spread corresponding to a leader's decision is estimated using the SAA method (see Algorithm \ref{SAA}) with a certain optimality gap. Hence, the optimality gaps attained basically determine the quality of the final solution. We assess the performance of the implemented SAA method by two measures: the estimate of the optimality gap and its variance. To this end, 10 instances are generated based on the Watts-Strogatz model, which produces networks with small mean distances between node pairs and a relatively high clustering coefficient, i.e, a small-world network reflecting the general structure of social networks better than completely random graphs. The number of nodes ($n$) in the generated networks take five different values, and the density of the number of arcs have two different values. The density is defined as the ratio of the number of arcs existing in the network to the number of arcs found in a complete network, namely $d=\frac{m}{n(n-1)}$.

The arc weights $w_{ij}$ are generated uniformly in the unit interval $[0,1]$, and then normalized such that $\sum_{j \in V} w_{ji} \leq 1$. This allows the possibility that a node cannot be influenced even if all of its preceding neighbors (nodes that are the tails of arcs connecting them with the node in question) are influenced, which is also the case in the original LT model. The leader's solutions, i.e., seed sets are generated randomly for the fixed seed size of $\lfloor 0.15 \, n\rfloor$, since the goal of these experiments is to analyze the quality of the solutions obtained for the LLP.

The first part of the experiments is devoted to the cardinality-based deactivation where the deactivation costs are assumed to be unity (i.e., $e_i=1$ for all $i$). This implies that the deactivation budget $E$ represents the number of nodes in the seed set that can be deactivated. In the second part of the experiments, we use cost-based deactivation in which deactivation costs $e_i$ are generated uniformly from the set $\{10,15,20\}$ corresponding to low, medium, and high unit deactivation cost levels.

The threshold distribution follows a continuous uniform distribution in the interval $[0,1]$. The threshold samples used in Algorithm \ref{SAA} are generated using Latin Hypercube Sampling method since it was shown to produce smaller optimality gap and variance estimates compared to simple random sampling \citep{stein1987large}. In the first stage of the SAA method, ALLP is solved $M=20$ times using samples of size $N=50$ by means of CPLEX 12.7 for the selected leader's solutions. In the second stage, the evaluation sample size $N'$ is set to 2000 and in stage 3, the re-evaluation sample size $N''$ is taken as 10,000.

The results are presented in Table \ref{table:Follower Response-equal} and Table \ref{table:Follower Response-unequal}. In each table, the first column provides the instance characteristics including the network size $n$, the number of arcs $m$, and the cardinality of the leader's seed set $C$. The deactivation budget $E$ is shown in the second column. The upper bound (UB) represents the objective value $\hat{z}_{N''}(\mathbf{x},\tilde{\mathbf{y}}^*)$ that Algorithm \ref{SAA} yields, and the lower bound (LB) stands for $\bar{z}_N^M (\mathbf{x})$ which is the output of the first stage of Algorithm \ref{SAA}. The remaining columns denote the estimate of the percent optimality gap, a $95\%$ confidence interval (CI) on the true gap as a percent of the UB, and the solution time, respectively. Notice that the confidence interval computed by the expression given in \eqref{CI} represents the variance of the optimality gap:
\begin{equation}
100\times \frac{\hat{z}_{N''}(\mathbf{x},\tilde{\mathbf{y}}^*) - \bar{z}_N^M(\mathbf{x}) \pm t_{\alpha/2, \nu}\sqrt{V(\hat{z}_{N''}(\mathbf{x},\tilde{\mathbf{y}}^*)) + V(\bar{z}_N^M(\mathbf{x}))} } {\hat{z}_{N''}(\mathbf{x},\tilde{\mathbf{y}}^*)}, \label{CI}
\end{equation}
where $\nu$ is the approximate degrees of freedom \citet{montgomery2010applied} given as
\begin{equation}
\nu=\frac{ \big( V(\hat{z}_{N''}(\mathbf{x},\tilde{\mathbf{y}}^{\star})) + V(\bar{z}_N^M(\mathbf{x})) \big)^2}
{\frac{V( \hat{z}_{N''}(\mathbf{x},\tilde{\mathbf{y}}^{\star}))^2} {N''-1} + \frac{V(\bar{z}_N^M(\mathbf{x}))^2}{M-1}}.
\end{equation}

It can be observed in Table \ref{table:Follower Response-equal} and Table \ref{table:Follower Response-unequal} that the negative optimality gap estimates are obtained for some instances. This is not unusual in the SAA method \citep{mak1999monte}. It can also be observed that as the network becomes denser, the optimality gap estimates become higher, and that the solution time increases rapidly with the number of arcs. According to the additional experimental results not reported here, for instances with $n$ ranging in the interval $[100-1000]$, the solution time of the ALLP solved in the first stage of Algorithm \ref{SAA} dominates the overall solution time as the network becomes larger (higher values of $n$) and denser (higher values of $m$). This is the reason why increasing the sample size ($N$) or the number of batches ($M$) to obtain better lower bound estimates is not practical.
\begin{table}[h!]
\caption{Solution of the LLP for leader strategy $\mathbf{x}$ under cardinality-based deactivation}
\vspace{-10pt}
	\begin{center}
		\begin{tabular}{c|ccccccc}
			\hline
        	&	    &		&		&	Optimality  	&	Confidence Interval 	&	Computation 	\\
$(n,m,C)$	&	$E$	&	UB	&	LB	&	Gap (\%)	    &	on the Gap (\%)      	&	Time (s)	\\	\hline
(20, 40, 3)	&	1	&	6.30	&	6.15	&	2.30	&	(2.18,2.42)	&	2	\\	
			&	2	&	2.54	&	2.55	&	-0.35	&	(-0.5,-0.21)&	2	\\	\hline
(20, 80, 3)	&	1	&	6.10	&	5.91	&	3.02	&	(2.62,3.42)	&	3	\\	
			&	2	&	2.78	&	2.81	&	-1.03	&	(-1.47,-0.6)&	4	\\	\hline
(40, 80, 6)	&	2	&	10.34	&	10.33	&	0.04	&	(-0.32,0.41)&	4	\\	
			&	3	&	5.33	&	5.37	&	-0.67	&	(-0.97,-0.36)&	5	\\	\hline
(40, 160, 6)&	2	&	14.48	&	14.15	&	2.26	&	(1.73,2.78)	&	13	\\	
			&	3	&	10.09	&	9.61	&	4.71	&	(4.11,5.31)	&	9	\\	\hline
(60, 120, 9)&	3	&	12.99	&	12.86	&	0.95	&	(0.77,1.13)	&	8	\\	
			&	4	&	10.42	&	10.37	&	0.51	&	(0.45,0.56)	&	7	\\	\hline
(60, 240, 9)&	3	&	22.25	&	21.84	&	1.85	&	(1.47,2.22)	&	16	\\	
			&	4	&	17.50	&	17.17	&	1.88	&	(1.24,2.52)	&	39	\\	\hline
(80, 160, 12)&	4	&	18.28	&	18.06	&	1.23	&	(1.09,1.38)	&	10	\\	
			&	6	&	9.47	&	9.45	&	0.20	&	(0.03,0.38)	&	17	\\	\hline
(80, 320, 12)&	4	&	32.44	&	31.65	&	2.42	&	(1.96,2.88)	&	45	\\	
			&	6	&	26.05	&	25.78	&	1.01	&	(0.35,1.68)	&	151 \\	\hline
(100, 200, 15)&	5	&	16.53	&	16.52	&	0.01	&	(-0.2,0.21)	&	13	\\	
			&	7	&	17.78	&	17.67	&	0.60	&	(0.46,0.75)	&	10	\\	\hline
(100, 400, 15)&	5	&	41.03	&	40.09	&	2.30	&	(1.35,3.25)	&	119	\\	
			&	7	&	30.64	&	30.35	&	0.94	&	(-0.29,2.16)&	130	\\	\hline
		\end{tabular}
\label{table:Follower Response-equal}
\end{center}
\end{table}

\begin{table}[h!]
\caption{Solution of the LLP for leader strategy $\mathbf{x}$ under cost-based deactivation}
\vspace{-10pt}
	\begin{center}
		\begin{tabular}{c|ccccccc}
			\hline
        	&	    &		&		&	Optimality  	&	Confidence Interval 	&	Computation 	\\
$(n,m,C)$	&	$E$	&	UB	&	LB	&	Gap (\%)      	&	CI on Gap (\%)       	&	Time (s)	\\	\hline
(20, 40, 3)	&	15	&	5.46	&	5.50	&	-0.75	&	(-0.88,-0.62)	&	2	\\	
	&	20	&	3.38	&	3.34	&	1.21	&	(1.02,1.40)	&	2	\\	\hline
(20, 80, 3)	&	15	&	8.79	&	8.61	&	1.97	&	(1.70,2.23)	&	2	\\	
	&	20	&	4.52	&	4.58	&	-1.25	&	(-1.41,-1.09)	&	3	\\	\hline
(40, 80, 6)	&	20	&	14.30	&	14.45	&	-1.06	&	(-1.17,-0.96)	&	4	\\	
	&	30	&	9.84	&	9.87	&	-0.31	&	(-0.48,-0.15)	&	4	\\	\hline
(40, 160, 6)	&	20	&	25.78	&	25.03	&	2.91	&	(2.55,3.26)	&	8	\\	
	&	30	&	17.03	&	16.56	&	2.77	&	(2.43,3.12)	&	7	\\	\hline
(60, 120, 9)	&	30	&	18.35	&	18.30	&	0.26	&	(0.11,0.41)	&	6	\\	
	&	45	&	17.25	&	17.23	&	0.10	&	(0.03,0.17)	&	6	\\	\hline
(60, 240, 9)	&	30	&	22.82	&	22.39	&	1.92	&	(1.13,2.72)	&	13	\\	
	&	45	&	17.68	&	17.04	&	3.63	&	(3.12,4.14)	&	34	\\	\hline
(80, 160, 12)	&	40	&	21.52	&	21.58	&	-0.27	&	(-0.44,-0.09)	&	8	\\	
	&	60	&	14.00	&	14.02	&	-0.17	&	(-0.37,0.03)	&	8	\\	\hline
(80, 320, 12)	&	40	&	37.69	&	37.75	&	-0.17	&	(-0.68,0.34)	&	67	\\	
	&	60	&	28.87	&	28.79	&	0.29	&	(-0.36,0.93)	&	52	\\	\hline
(100, 200, 15)	&	50	&	24.27	&	24.23	&	0.15	&	(0.02,0.27)	&	11	\\	
	&	75	&	18.37	&	18.34	&	0.12	&	(-0.01,0.25)	&	10	\\	\hline
(100, 400, 15)	&	50	&	51.13	&	50.05	&	2.11	&	(1.19,3.02)	&	193	\\	
	&	75	&	33.44	&	31.83	&	4.80	&	(3.96,5.64)	&	98	\\	\hline

		\end{tabular}
\label{table:Follower Response-unequal}
\end{center}
\end{table}

\subsection{Assessment of the Performances of SAM and TSM}
Convinced with the quality of results provided by the SAA method, we focus now on the assessment of the performances of the proposed matheuristics SAM and TSM. We first consider small test instances that can also be solved by complete enumeration. Small-world networks are generated for two different network sizes with $n=20,30$ and two density values $d=0.1,0.2$. For each combination, five distinct network structures are considered, which results in 20 instances. Each instance is solved three times, and the results are displayed in Table \ref{table:Heuristic vs complete-equal} for cardinality-based IMPD instances and Table \ref{table:Heuristic vs complete-unequal} for cost-based IMPD instances on the basis of averages over three replications. For the cardinality-based IMPD instances, the seed selection and deactivation costs are unity. Hence $C$ represents the number of nodes selected as the seed, and $E$ designates the number of deactivated nodes. For the cost-based IMPD instances, the activation costs $c_i$ and the deactivation costs $e_i$ are generated randomly from the set $\{10,15,20\}$, independent from each other.

Two approaches were used for the generation of the initial solutions as explained before. We found out that for the cost-based IMPD instances, the second approach that uses only the costs yields better results. For the cardinality-based IMPD instances, the initial solution method takes into account only the individual spread values since the costs are equal in this type of instances.

In all of our experiments, the same parameter values given in the previous subsection are used for the SAA algorithm. The $t_{\max}$ parameter is set to three hours and the best solution found is reported hourly in both matheuristics. Moreover, each of the three move types is equally likely to be selected in both SAM and TSM for the cost-based IMPD instances. Only 1-Swap move is implemented for SAM and TSM when cardinality-based IMPD instances are solved. The remaining parameter values in SAM are set as follows: $p_{0}=0.8$, $r=0.9$, $\gamma =0.2$, and $\phi=0.5$. In TSM, the proportion of the neighboring solutions examined, or equivalently the candidate list size of TSM is set to $\tau=0.5$. The penalty coefficient $\mu$ for the long-term frequency-based memory is taken equal to one.

The first column in both tables show the network size in terms of $n$ and $m$, as well as the activation and deactivation budget levels $C$ and $E$. Please note that budget levels are determined in such a way that the solution time of complete enumeration is acceptable. In the next two columns, the average optimality gaps of the best solutions found by SAM and TSM are provided as a percentage of the optimal objective value $\hat{z}_L$ associated with each instance.

\begin{table}[h!]
\caption{Assessment of the results obtained by SAM and TSM for cardinality-based IMPD instances}
\vspace{-10pt}
	\begin{center}
		\begin{tabular}{c|cccc}
			\hline
            &              &              &  Compl. Enum.     &   Compl. Enum. \\
$(n,m,C,E$) & SAM Gap (\%) & TSM Gap (\%) & $\hat{z}_L$         & CPU Time (s) \\ \hline
	&	0.00	&	0.00	&	8.84	&	2898	\\	
	&	0.00	&	0.00	&	9.05	&	2789	\\	
$(20,40,3,1)$	&	0.00	&	0.00	&	8.76	&	2715	\\	
	&	0.00	&	0.00	&	9.91	&	2521	\\	
	&	0.00	&	0.00	&	6.54	&	2494	\\	\hline
	&	0.00	&	0.00	&	10.81	&	3563	\\	
	&	0.00	&	0.00	&	10.60	&	3619	\\	
$(20,80,3,1)$	&	0.00	&	0.00	&	10.11	&	3514	\\	
	&	0.00	&	0.00	&	10.72	&	3860	\\	
	&	1.61	&	0.00	&	11.67	&	3603	\\	\hline
Average	&	0.16	&	0.00	&		&	3186	\\	\hline
	&	0.05	&	0.00	&	11.89	&	112,594	\\	
	&	1.10	&	0.00	&	12.30	&	126,877	\\	
$(30,90,4,2)$	&	0.00	&	0.00	&	11.75	&	125,100	\\	
	&	0.69	&	0.00	&	12.14	&	129,783	\\	
	&	0.00	&	0.00	&	13.59	&	122,138	\\	\hline
	&	0.00	&	0.00	&	13.84	&	143,575	\\	
	&	0.00	&	0.00	&	12.82	&	135,826	\\	
$(30,180,4,2)$	&	0.00	&	0.00	&	12.22	&	153,187	\\	
	&	0.00	&	0.00	&	12.95	&	155,080	\\	
	&	0.21	&	0.31	&	12.36	&	168,366	\\	\hline
Average	&	0.20	&	0.03	&		&	139,992	\\	\hline
		\end{tabular}
\label{table:Heuristic vs complete-equal}
\end{center}
\end{table}

\begin{table}[h!]
\caption{Assessment of the results obtained by SAM and TSM for cost-based IMPD instances}
\vspace{-10pt}
	\begin{center}
		\begin{tabular}{c|cccc}
			\hline
            &              &              &  Compl. Enum.     &   Compl. Enum. \\
$(n,m,C,E$) & SAM Gap (\%) & TSM Gap (\%) & $\hat{z}_L$         & CPU Time (s) \\ \hline
	&	0.00	&	0.00	&	8.39	&	1287	\\	
	&	0.00	&	0.00	&	8.26	&	901	\\	
$(20,40,40,20)$	&	2.33	&	0.00	&	9.43	&	1709	\\	
	&	0.00	&	1.78	&	9.51	&	1063	\\	
	&	0.54	&	3.52	&	7.94	&	1622	\\	\hline
	&	0.00	&	0.00	&	12.37	&	1538	\\	
	&	0.00	&	0.00	&	11.08	&	2255	\\	
$(20,80,40,20)$	&	0.00	&	1.26	&	10.35	&	1943	\\	
	&	1.97	&	3.56	&	11.82	&	1727	\\	
	&	2.07	&	0.00	&	11.09	&	1479	\\	\hline
	Average	&	0.69	&	1.01	&		&	1552\\	\hline

	&	2.01	&	0.00	&	16.66	&	86,299	\\
	&	0.74	&	0.74	&	18.88	&	84,469	\\
$(30,90,60,30)$	&	1.34	&	1.34	&	19.72	&	89,195	\\
	&	1.52	&	0.00	&	18.67	&	120,759	\\
	&	6.22	&	0.00	&	15.24	&	136,490	\\ \hline
	&	0.68	&	0.00	&	21.57	&	121,654	\\
	&	0.00	&	0.00	&	16.68	&	108,353	\\
$(30,180,60,30)$	&	0.97	&	0.00	&	16.59	&	111,755	\\
	&	1.84	&	0.00	&	18.73	&	108,033	\\
	&	0.00	&	0.00	&	17.67	&	127,630	\\ \hline
Average &		1.53	&	0.21		&&		109,464 \\	\hline
		\end{tabular}
\label{table:Heuristic vs complete-unequal}
\end{center}
\end{table}

On the basis of the optimality gaps given in both tables, we can observe that they are quite satisfactory. Especially in the case of cardinality-based IMPD instances, using only 1-Swap moves seems to indicate a better performance in terms of searching the solution space. In addition, TSM yields much smaller optimality gaps within the time limit. One possible reason may be that the actual running time of SAM is less than three hours on the average, because it terminates before $t_{\max}$ when it gets stuck at a local optimal solution. To analyze this issue, we compare the hourly optimality gaps of instances with $n=30$, that are presented in Table \ref{table:Hourly results-both}. The third-to-fifth columns are the end-of-hour average optimality gaps over 10 instances and three replications. The last column gives the average running time of the algorithms. It can be seen that SAM terminates before TSM in most of the instances, especially in the cost-based instances. However, TSM outperforms SAM at all epochs, even at the end of the first hour in all instance types.

\begin{table}[h!]
\caption{Results obtained by SAM and TSM }
\vspace{-10pt}
	\begin{center}
		\begin{tabular}{lc|cccc} \hline
Instance           	&	Matheuristic   &   $\text{Gap}_1$	&	$\text{Gap}_2$	&	Final Gap &   CPU Time (s)	\\	\hline
Cardinality-based   &      SAM         &	1.97	            &	0.42             	&	0.20	  &	      9269	    \\	
Cardinality-based   &      TSM	       &	0.18	            &	0.13             	&	0.03	  &	     10,962	\\	\hline
Cost-based          &      SAM         &	3.43             	&	1.59	            &	1.53	  &	     6387   	\\	
Cost-based          &      TSM	       &	1.36	            &	0.51	            &	0.21	  &	     10,887	\\	\hline
\end{tabular}
\label{table:Hourly results-both}
\end{center}
\end{table}

\subsection{Comparison of SAM and TSM for larger instances}
In this section, we compare the performances of our matheuristics on IMPD instances the optimal objective values of which cannot be attained by complete enumeration due to excessive solution times. To this end, we utilize the cost-based instances that were used in Section \ref{subsection:SAA Method Results}. The activation budget values are determined in such a way that the leader can select at most 10\% of the nodes as the seed set in the network and the follower can deactivate at most half of the nodes in the seed set. The SAM objective values are used as the baseline and the relative performance of TSM is reported as $\Delta=100\times(\hat{z}_{TSM}^{*}-\hat{z}_{SAM}^{*})/\hat{z}_{SAM}^{*}$. Here, $\hat{z}_{SAM}^{*}$ ($\hat{z}_{TSM}^{*}$) denotes the average objective value that SAM (TSM) yields over three replications. A positive $\Delta$ value indicates that TSM finds better solutions on the average. In Table \ref{table:SA vs TS}, we report the results by presenting the end-of-hour performances. The $t_{\max}$ parameter is set to three hours except for the instances $(80,320)$ and $(100,400)$ which require significantly larger solution times by the SAA method compared to others. For these two instances, $t_{\max}$ is set to six hours and $\Delta_1$, $\Delta_2$, and $\Delta_3$ correspond to the results at the end of $3^\text{rd}$, $4^\text{th}$ and $6^\text{th}$ hours. The results show that TSM outperforms SAM in most of the instances, even though its superiority decreases in time.

\begin{table}[h!]
\caption{Comparison SAM and TSM algorithms}
\vspace{-10pt}
	\begin{center}
		\begin{tabular}{c|ccc}
		\hline
$(n,m)$ & $\Delta_1$&$\Delta_2$&$\Delta_3$ \\ \hline
$(20,40)$	&	0.00	&	0.00	&	0.00	\\
$(20,80)$	&	-2.95	&	-2.95	&	-2.95	\\
$(40,80)$	&	-3.65	&	-1.76	&	0.13	\\
$(40,160)$	&	8.52	&	2.64	&	0.00	\\
$(60,120)$	&	0.51	&	-4.76	&	-3.22	\\
$(60,240)$	&	5.46	&	2.50	&	0.27	\\
$(80,160)$	&	34.60	&	6.23	&	0.98	\\
$(80,320)$	&	24.33	&	35.32	&	24.59	\\
$(100,200)$	&	21.58	&	23.43	&	8.04	\\
$(100,400)$	&	8.48	&	8.88	&	8.88	\\ \hline
Average	&	9.69	&	6.95	&	3.67	\\ \hline
\end{tabular}
\label{table:SA vs TS}
\end{center}
\end{table}

\subsection{TSM results on a real social network}
\citet{newman2001structure} investigated various scientific collaboration networks including co-authorship networks in different disciplines. All these instances have the small-world property and a high clustering coefficient. Since these are the key features of social networks, co-authorship datasets have frequently been used in many social network analysis studies such as \citet{Kempe}. The final experiments in our paper are conducted on the co-authorship network obtained from the physics section of \texttt{www.arXiv.org}. The authors are represented by nodes, and each paper is represented by an arc (from the first author to other co-authors). The original dataset has 37,154 nodes and 231,507 arcs. When parallel arcs between two nodes are removed, the resulting number of arcs is 180,826 and the average out-degree is 4.87. This network is quite large to implement the matheuristic methods developed for the solution of the bilevel IMPD. However, smaller networks can be extracted by selecting a subset of nodes and arcs from this network. When a node subset of size $n=100$ together with the arcs among these nodes is selected randomly, the resulting network might have no arcs at all or can be very sparse. Similarly, when a set of arcs along with the nodes incident to them is selected randomly, the resulting graph looks almost like a matching. The optimal solution on such a network is trivial, where the optimal spread is $2(C-E)$ for a cardinality-based IMPD instance.

Hence, rather than randomly taking nodes into the network, we follow the following approach. Two networks of size $n=100$ and $n=200$ are generated by selecting nodes with a large out-degree so as to obtain a similar network to the original one in terms of the average out-degree. By focusing on cardinality-based IMPD instances with unit seed selection and deactivation costs and applying only TSM due to its higher performance on previous experiments, we solve a number of instances. The initial solution consists of the nodes with the largest out-degree. TSM is executed for 24 hours, and the results are reported in Table \ref{table:Arxiv Results}, where $\hat{z}_t$ represents the best objective value found at the end of $t$ hours. We remark that the average out-degree for $n=100$ and $n=200$ instances are 4.18 and 4.33, respectively. Recall that $\tau$ is the proportion of solutions evaluated in the neighborhood of the current solution at each iteration of TSM.

\begin{table}[h!]
  \centering
  \caption{Results of TSM on instances based on ArXiv Hep-Th dataset}
    \begin{tabular}{c|c|c|c|cccccc}
\hline
    $n$ & $\tau$ & $C$ & $E$ & $\hat{z}_1$ & $\hat{z}_2$ & $\hat{z}_4$ & $\hat{z}_8$ & $\hat{z}_{12}$ & $\hat{z}_{\mathrm{Final}}$ \\
\hline
    \multirow{8}[2]{*}{100} & 0.05  & 5     & 1     & 27.46 & 27.46 & 30.80 & 30.80 & 30.80 & 30.80 \\
          & 0.10  & 5     & 1     & 27.46 & 32.18 & 32.18 & 32.18 & 32.18 & 32.18 \\
          & 0.05  & 5     & 2     & 18.85 & 18.85 & 22.19 & 22.19 & 22.19 & 22.44 \\
          & 0.10  & 5     & 2     & 18.85 & 23.48 & 23.48 & 23.48 & 23.48 & 23.48 \\
          & 0.05  & 10    & 2     & 33.24 & 39.20 & 40.67 & 47.87 & 49.02 & 49.02 \\
          & 0.10  & 10    & 2     & 30.22 & 37.84 & 40.67 & 47.87 & 49.02 & 49.02 \\
          & 0.05  & 10    & 4     & 19.40 & 21.52 & 22.46 & 28.76 & 32.51 & 32.51 \\
          & 0.10  & 10    & 4     & 16.84 & 19.40 & 21.52 & 24.87 & 27.26 & 30.26 \\
\hline
    \multirow{4}[2]{*}{200} & 0.05  & 10    & 2     & 29.73 & 29.73 & 34.61 & 47.91 & 62.05 & 71.95 \\
          & 0.05  & 10    & 4     & 20.00 & 20.00 & 20.00 & 24.89 & 24.89 & 31.92 \\
          & 0.05  & 20    & 4     & 40.14 & 40.14 & 40.14 & 46.11 & 46.11 & 62.98 \\
          & 0.05  & 20    & 8     & 25.89 & 25.89 & 25.89 & 25.89 & 25.89 & 28.79 \\
\hline
    \end{tabular}%
  \label{table:Arxiv Results}%
\end{table}%

Observe that a higher value of $\tau$ does not always lead to a better solution as can be seen from the results of the instances with $n=100$. This is a fortunate outcome since the  solution times per iteration are excessive for $\tau=0.10$ when solving instances with $n=200$ nodes. Thus, we only implement TSM with $\tau=0.05$ for these instances.

\section{Conclusion} \label{section:Conclusion}
This work deals with a competitive Influence Maximization Problem which can be formulated as a Stackelberg game. There are two players who make decisions sequentially. The first player (leader) wants to maximize the spread by activating an influential seed set, and the second player (follower) tries to minimize it by deactivating some of the activated nodes. This problem, referred to as Influence Maximization Problem with Deactivation, is formulated as a bilevel integer programming model. The lower level problem is an integer stochastic optimization problem as a consequence of the uncertainty in the node threshold values. This choice of the threshold values correspond to the well-known linear threshold diffusion model which, along with the independent cascade model, is one of the most widely adopted diffusion models. Sample Average Approximation method is implemented in order to approximate the follower's optimal objective value in the lower level problem. The best decisions of the leader are searched for in the upper level problem by two metaheuristics based on simulated annealing and tabu search, respectively. The performances of the resulting matheuristics, called SAM and TSM, are compared with the complete enumeration method on small test instances, while they are compared against each other on larger instances.

The solution times of SAM and TSM are dominated by the Sample Average Approximation method. Therefore, a possible future research direction can be the improvement of the solution approach of the follower's lower level problem in terms of efficiency. For example, a heuristic method can be employed instead of a mathematical model. This would enable us to explore the solution space of the leader more efficiently, which ultimately can lead to tackle larger problem instances.

\section*{Acknowledgements}
This work is partially supported by Bo\u{g}azi\c{c}i University Scientific Research Project under the Grant number: BAP 12745.

\bibliographystyle{apalike}
\bibliography{mybib}

\end{document}